\documentclass[prd,nofootinbib,aps,floats,floatfix,amssymb,secnumarabic,twocolumn]{revtex4-1} %
\usepackage{amsmath}
\usepackage{bm}
\usepackage{graphicx}
\usepackage[utf8]{inputenc}
\usepackage{color}
\usepackage{hyperref}
\usepackage{cancel}
\usepackage{slashed}
\newcommand{\be}{\begin{equation}}
\newcommand{\ee}{\end{equation}}
\def\bea{\begin{eqnarray}}
\def\eea{\end{eqnarray}}
\newcommand{\sss}{\scriptscriptstyle}

\newcommand{\nn}{\nonumber}
\newcommand{\ua}{\uparrow}
\newcommand{\da}{\downarrow}
\newcommand{\W}{{\sss W}}
\newcommand{\Z}{{\sss Z}}
\newcommand{\h}{{h}}
\newcommand{\f}{{f}}
\newcommand{\HC}{{\sss\rm HC}}
\newcommand{\HB}{{\sss\rm HB}}

\newcommand{\jc}[1]{{{#1}}}

\begin{document}
\title{Challenges for models with composite states}
\author{James M.\ Cline\footnote{jcline@physics.mcgill.ca}}
\affiliation{Department of Physics, McGill University,
3600 Rue University, Montr\'eal, Qu\'ebec, Canada H3A 2T8}
\author{Weicong Huang\footnote{huangwc@itp.ac.cn}}
\affiliation{Key Laboratory of Theoretical Physics, Institute of Theoretical Physics, Chinese Academy of Sciences, Beijing 100190, China and \\
Department of Physics, McGill University,
3600 Rue University, Montr\'eal, Qu\'ebec, Canada H3A 2T8}
\author{Guy D.\ Moore\footnote{guy.moore@physik.tu-darmstadt.de}}
\affiliation{Institut f\"ur Kernphysik, Technische Universit\"at Darmstadt
Schlossgartenstrasse 2, D-64289 Darmstadt, Germany}

\begin{abstract}

Composite states of electrically charged and QCD-colored hyperquarks
(HQs) in a confining SU($N_{\HC}$) hypercolor gauge sector are a plausible
extension of the standard model at the TeV scale, and have been widely
considered as an explanation for the tentative LHC diphoton excess. 
Additional new physics is required to avoid a stable charged
hyperbaryon in such theories.  We classify renormalizable models
allowing the decay of this unwanted relic  directly into standard
model states, showing that they  are significantly restricted if the
new scalar states needed for UV completion are at the TeV scale.
Alternatively, if hyperbaryon number is conserved, the charged relic
can decay into a neutral hyperbaryon.  Such theories are strongly
constrained by direct detection, if the neutral constituent hyperquark
carries color or weak isospin, and by LHC searches for  leptoquarks 
if it is a color singlet.  We show that the neutral hyperbaryon can
have the observed relic abundance if the confinement scale 
and the hyperquark mass are above TeV scale, even in the absence
of any hyperbaryon asymmetry.

\end{abstract}
\maketitle

\section{Introduction}
\label{intro}

Hints of a small excess of events in  the diphoton channel have been
reported by  ATLAS 
\cite{ATLAS:diphoton,Aaboud:2016tru} and CMS
\cite{CMS:2015dxe,Khachatryan:2016hje} experiments during the 13 TeV run.  Although they
are probably a statistical fluctuation, it is intriguing that both
experiments see the excess at the same invariant mass of the photon
pair, at approximately $m_{\gamma\gamma} = 750\,$GeV, and that there
are further hints of an excess at higher invariant masses
\cite{Aaboud:2016tru}.  This has
prompted numerous theoretical interpretations in terms of a spin-0
resonance decaying into photons.  A plausible class of models
considers the resonance to be composed of heavy constituents $\Psi$ which we
will call hyperquarks (HQs), bound by an SU($N_{\HC}$) hypercolor
confining gauge theory.  If the HQs carry electric charge then the
pion- or quarkonium-like bound state assumed to be the 750 GeV
resonance can decay into photons to explain the observed 
signal \cite{Harigaya:2015ezk,Nakai:2015ptz,Redi:2016kip,Craig:2015lra,Bian:2015kjt,Bai:2015nbs,
Harigaya:2016pnu,Harigaya:2016eol,Kamenik:2016izk,Ko:2016sht,
Foot:2016llc,Bai:2016czm,Bai:2016vca}. 

This kind of extension of the standard model seems natural since it
simply enlarges the gauge symmetry group by an additional
SU($N_{\HC}$)  factor.  One might expect that, similarly to the
standard model, the HQs carry a new, possibly conserved charge,
hyperbaryon (HB) number.  In this work we explore the consequences of
HB number being conserved, leading to a dark matter candidate but also
potentially severe conflicts with observation.\footnote{For a
discussion of more exotic hyperbaryons, consisting of bound states of
hyperquarks in several different representations of the standard model gauge
symmetries, see ref.\ \cite{Antipin:2015xia}} \ \ We also consider
models in which it is broken by renormalizable interactions, which
turn out to be
more constrained than one might at first think.  Although the tentative
excess of 750 GeV diphotons at LHC motivated our study, it could be of
more general interest even if this signal does not persist, since it
has now been established that new physics of this kind could be on 
the verge of discovery at LHC.  In the following we will focus on
models that predict a  750 GeV bound state $\tilde\pi$ that decays into
photons, but our observations could obviously be adapted to other similar
models.

We will assume that $\Psi$ is vectorlike.  If
$m_\Psi \ll \Lambda_{\HC}$ then the bound state $\tilde\pi$ is pion-like,
with a mass
scaling as  $m_{\tilde\pi}\sim \sqrt{m_\Psi\Lambda_{\HC}}$ due to an assumed
approximate chiral symmetry, softly broken by $m_\Psi$.  If
$m_\Psi >\Lambda_{\HC}$, the composite state would be more similar to
charmonium.  However the fact that the putative $750\,$GeV resonance
is relatively narrow and distinct indicates that $m_\Psi$ cannot be
much greater than $\Lambda_\HC$; otherwise one would expect to produce a
series of closely-spaced resonances with fractional mass splitting
$\Delta m/m\sim (\Lambda_{\HC}/m_\Psi)^{3/4}$, based upon a semiclassical
model of bound states in a linear confining potential 
$V\sim \Lambda_{\HC}^2r$.

In these models, it is assumed that the HQs are also colored and can
thus be produced by gluon-gluon fusion ($gg$F).  Alternatively, it is possible to
have sufficient production through photon-photon or vector boson
fusion 
\cite{Fichet:2015vvy,Csaki:2015vek,Harland-Lang:2016qjy,
Iwamoto:2016ral,Appelquist:2016pel}
if the HQ carries a large hypercharge $\sim (3-4)$, indicating a
Landau pole at a relatively low scale, barring additional
states.   However the  modest growth of the photon parton
distribution function (PDF) with energy puts this scenario in tension
with the lack of any observed signal in the 8 TeV LHC
run.\footnote{However ref.\ \cite{Harland-Lang:2016qjy} points out that this is subject
to uncertainties in the parton distribution functions and finds that
the tension is not strong.}
For these reasons
models with $gg$F production are favored, and our focus will therefore
be on HQs that carry QCD color.  We will comment on colorless models
in section \ref{nocolor}.

One possibility is that there is a conserved HQ number that 
leads to a stable hyperbaryon (HB) consisting of $N_{\HC}$ HQs.  There
are very stringent constraints on electrically charged relics, so that
a realistic model should provide some way for these unwanted relics to
decay.  In some cases it is possible to write down a high-dimensional
effective operator that would allow the charged HB to decay directly
into standard model (SM) particles.  However it is theoretically more
satisfying to demonstrate the renormalizable interactions that would
allow for the decays of the HB, either into purely SM particles, or
into an electrically neutral HB that might be a viable dark matter
candidate.  One point of the present work is that there
are relatively few categories of renormalizable models that lead to
nonconserved HB number, and they are strongly constrained by 
collider searches if the new scalars that must be added are at the
TeV scale.  We survey these possibilities in section \ref{theorem}.

If HB number is conserved, the lightest HB must be electrically
neutral, and is a dark matter candidate.  {\it A priori}, 
it could carry SU(3)$_c$ color, in which case it binds to ordinary
quarks or antiquarks to make a color-neutral composite state, whose
residual strong interactions give it a large cross
section for scattering on nucleons.  This is \jc{likely to be
excluded at the same level as charged relics by searches for anomalous heavy isotopes}, favoring models in which the lightest
stable HQ is a color singlet.  If it is an SU(2)$_L$ doublet, the
constraints \jc{from direct detection} are less severe, but still quite significant.  The safest
case with constituents that are purely neutral under SM interactions 
turns out to have a relatively light leptoquark bound state, assuming
that ordinary baryon number is still an accidental symmetry of the
full theory; hence even this case comes under pressure from current
LHC constraints.  These
scenarios are discussed in section \ref{HBcons}.

The relic density of conserved HBs could be due to an asymmetry,
analogous to the baryon asymmetry.  Here we suppose that the mechanism
for generating an HB asymmetry is weak or lacking, and focus on the
symmetric component, which should be understood in any case before
invoking an asymmetry.  In section \ref{relic} we compute the
abundance for purely singlet HBs and for those that carry weak
isospin, showing its dependence on the confinement scale
$\Lambda_{\HC}$ and the mass of the neutral HQ.  
Even the purely singlet HB has electromagnetic interactions
with protons through loops containing the charged HQ.  This leads to
weak constraints from direct detection that we derive in section 
\ref{EM}.  Models in which the charged HQ does not carry QCD color 
are much less restricted by the considerations of the previous 
sections.  We briefly comment on them in section \ref{nocolor},
and give conclusions in section \ref{conclusions}.

\section{Hyperbaryon number violation}
\label{theorem}

First we deal with the possibility that hyperbaryon number is
not a symmetry of the theory, and HBs can decay directly into
SM particles.  One might imagine further possibilities by allowing
new dark matter particles in the final states, but we do not pursue
this here.

Let us provisionally assume that the charged HQ  is a fermion
$\Psi_A^a$ with HC index $A$, color index $a$, and weak
hypercharge $Y$.   In addition, there may be bosonic fields carrying
fundamental HC indices. We can associate the global HB quantum number
1 to each  fundamental HC index, and $-1$ to each antifundamental
index. Thus a HC gauge boson, having one of each, has vanishing HB
number. If we were only allowed to contract HC indices in
fundamental/antifundamental pairs, then it would be impossible to
violate HB number while respecting the gauge symmetry.  However in
SU($N$) we also have the invariant tensor $\epsilon_{A_1,\dots,A_N}$.
This simple argument demonstrates that HB violation must involve
the $\epsilon$ tensor.  

We start by discussing a rather general class of models that lead to
decays of the hyperbaryons into standard model quarks.  Other types of
models have a structure depending upon the value of $N_{\HC}$, 
so we consider the possibilities $N_{\HC}=2,3,4$ in 
turn in the following.

\subsection{Neutral scalar hyperquarks}

There is a general class of models which have the same structure
and are renormalizable for $N_{\HC}=2,3,4$, requiring the presence
of $N_{\HC}$ flavors of fundamental SU($N_{\HC}$) scalars $\Phi_{i,A}$, and
that the hypercharge of $\Psi$ matches that of the SM $u_R$ or
$d_R$ quarks.  Illustrating the former case where $Y_\Psi=2/3$, 
it has the form
\be
	\sum_{i=1,j}\lambda_{ij}\bar\Psi^A_a\, \Phi_{i,A} u_{R,j}^a
	+\mu \, \epsilon^{A_1,\dots,A_{N}}\, \Phi_{1,A_1}\,\cdots
	\Phi_{N,A_{N}}
\label{Y23}
\ee
where $\mu$ has dimensions of (mass)$^{4-N}$ and $N=N_{\HC}$.  If the $\Phi$'s are heavy
they can be integrated out giving an operator schematically of the form
\be
	\frac{\mu\lambda^{N}}{m_\Phi^{2N}}\left(\bar\Psi u_R\right)^N
\label{HBdecay}
\ee
that allows the charged HB to decay into $N$ up-type quarks.  

On the
other hand if the $\Phi$'s are lighter than $m_\Psi$, then the
lightest HB is a scalar bound state $\Phi^N$, which can decay via
the operator $\mu\Phi^N$.  Concretely, this would allow the HB
to decay into $N-1$ hypermesons, 
\be
	\Phi^N\to (N-1)\,\Phi^*\Phi
\ee
by converting one $\Phi$ into $N-1$ 
antihyperquarks (or fewer, if final state $\bar\Phi$ annihilate with
$\Phi$ before hadronizing). \jc{ If $N\ge 4$ there is generically enough
phase space for the decay, even if the mases are dominated by the
constituents, since $N m_\Phi \ge 4 m_\Phi$.  For $N=3$
one would require the mesons to be pseudo-Goldstone bosons in order
to overcome this restriction. 
(For $N_{\HC}=2$ there is no clear distinction between a meson and
  a baryon since the fundamental representation of SU(2) is
pseudoreal.)}

\begin{figure}[t]
\centering
\centerline{\includegraphics[width=0.5\textwidth]{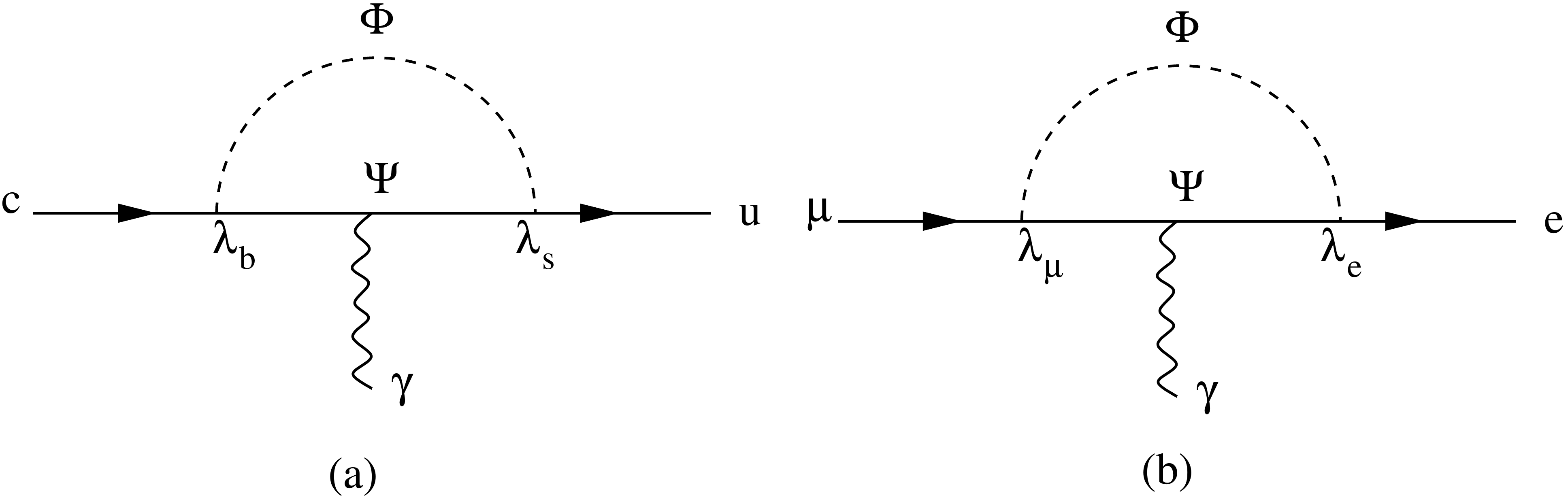}}
\caption{(a) Contribution to $c\to u\gamma$ from the model
eq.\ (\ref{Y23}).  (b) Contribution to $\mu\to e\gamma$ in the HB-violating
model of eq.\ (\ref{uvmodel}). }
\label{mu2egamma}
\end{figure}

\subsubsection{LHC constraints}
\label{lhcconst}

If $\Phi$ is heavy, then in addition to the HB-violating operator
(\ref{HBdecay}), there is a dimension-6 HB-conserving interaction
of the form
\be
	\frac{\lambda^2}{m_\Phi^2}\left|\bar\Psi u_R\right|^2
\label{piuu}
\ee
Because of its chiral structure it would
allow vector hypermesons to decay into $u_R$ quarks plus a gluon,
leading to three jets.\footnote{\jc{A different dimension-6 operator of
the form $(\bar\Psi\gamma_\mu\Psi)(\bar u_R\gamma^\mu u_R)$ would 
allow decays into two jets, but this operator is not induced by
the heavy scalar $\Phi$.}}\ \   However there is no obstruction to
making $\lambda^2/m_\Phi^2$ sufficiently small so that the branching
ratio for these decays is unimportant.  A lower bound (depending upon 
$N_{\HC}$ and $\mu$) can be placed on $\lambda/m_\Phi^2$ by demanding
that the charged HB decays before big bang nucleosynthesis (BBN), leading 
to  $\lambda/m_\Phi^2\gtrsim 10^{-4}/{\rm TeV}^2$ for the most
restrictive case of $N_{\HC}=4$, consistent with a small coefficient 
$\sim 10^{-8}/{\rm TeV}^2$ for the effective operator (\ref{piuu}),
if $m_\Phi \sim $TeV.

The case of light $\Phi$ is more interesting, since it leads to mass mixing between the SM quarks
and the composite fermions $U_i \equiv \bar\Psi\Phi_i$ that have the same quantum
numbers as $u_R$ (or $d_R$ in the alternate $Y_\Psi = -1/3$ models). 
The mixing comes from the Yukawa coupling 
$\lambda_{ij}\bar\Psi^A_a\, \Phi_{i,A} u_{R,j}^a$ leading to an
off-diagonal mass term of order 
$\lambda_{ij}\,f_{\tilde\pi}\, \bar U_i u_j$,
where $f_{\tilde\pi}$ is the hypermeson decay constant.

The heavy composite $U_i$ particles are constrained by LHC searches 
for heavy quarks.  
There will be Drell-Yan pair production of $\bar U_i U_i$, followed by
decays $U_i\to u_j h$ where $h$ is the Higgs boson, or
$U_i\to Wd_j$.  The first decay comes from mixing of $U_R$ with $u_R$,
while the second is due to $U_L$-$u_L$ mixing.  We note that
$U$ is a Dirac fermion since we have implicitly assumed that 
$\Psi$ is vectorlike in order to have a bare mass.  The mass matrix
takes the form
\be
	(\bar u_R\ \bar U_R)\left({m_u\atop 0}
	{\lambda f_{\tilde\pi}\atop M_U}\right)
	\left({u_L\atop U_L}\right)
\ee
where $m_u$ is the SM quark mass matrix and 
$M_U\sim \delta_{ij}\sqrt{\Lambda_{\HC}\,m_\Psi}\lesssim 750\,$GeV is the
mass matrix of the composite states in the absence of $U$-$u$ mixing.
After diagonalization, the left- and right-handed states have mixing
angles
\be
	\theta_L \sim {\lambda f_{\tilde\pi} m_u\over M_U^2},\quad
	\theta_R \sim {\lambda f_{\tilde\pi}\over M_U}
\ee
The effective couplings for $U\to hu$ and $U\to Wd$ are thus of
order
\be
	{m_u\over v} \theta_R = {\lambda\Lambda m_u\over v M_U},
	\quad g_2\theta_L = {g_2\lambda\Lambda m_u\over M_U^2}
\ee
implying that $U\to hu$ is the dominant decay channel, as long as
$m_U\gtrsim 300\,$GeV.  

This scenario has been considered by
ATLAS \cite{top-partner} and CMS \cite{Khachatryan:2015oba} (see also \cite{Aad:2015tba}) 
for the case of 
top partners decaying as $T\to ht$ and $Wb$.  These searches 
constrain
$m_T> 700-900\,$GeV depending upon the respective branching ratios,
with the strongest limit when $T\to ht$ dominates, as we expect here.
This contradicts the assumption that $m_\Phi< m_\Psi$ which would 
imply that $m_T < 750\,$GeV, hence ruling out a dominant coupling
to top quarks.

\subsubsection{Flavor constraints}
The interaction (\ref{Y23}) induces flavor changing neutral current
decays $c\to u\gamma$ as shown in fig.\ \ref{mu2egamma}.  The
analogous diagram gives $b\to s\gamma$ in the related model with
$Y_\Psi = -1/3$.   Defining
$t=m_\Psi^2/m_\Phi^2$ and writing the transition amplitude 
as $(m_b/\Lambda_b^2)\bar s_R \sigma^{\mu\nu}q_\nu\epsilon_\mu
b_L$, we find \cite{Lavoura:2003xp}
\be
	{1\over\Lambda_b^2} = {N_{\HC}(e/3)\lambda_b\lambda_s\over 
	32\pi^2 m_\Psi^2}
	f(t) < \left(1\over 55{\rm\ TeV}\right)^2
\label{b2sgamma}
\ee
where $f(t) = t/(t-1)^4[(t-1)(t^2-5t-2)/6 + t\ln t]\sim 0.1$ for
a typical value $t\sim 1.5$ and the experimental upper limit
is inferred from ref.\ \cite{Descotes-Genon:2015uva} (specifically,
$\Lambda_b^{-2} = 4\sqrt{2}\,G_F V_{tb} V^*_{ts} C'_7\, e/16\pi^2$ with
$C'_7 < 0.065$).  Taking
for example $m_\Psi=400\,$GeV  and $N_{\HC}=3$ we obtain a weak
constraint, $\sqrt{|\lambda_b\lambda_s|} 
\lesssim 0.75$.

\subsection{$N_{\HC}=2$}
We turn next to models that are specific to the value of 
$N_{\HC}$.  For $N_{\HC}=2$ we can construct the HB-violating
dimension-6 operators
\be
	\epsilon_{abc}\,\epsilon_{AB}(\bar\Psi^c_{A,a}\Psi_{B,b})
	(\bar u_{R,c}^c l_R,\quad \bar L_L^c Q_{L,c},\quad 
	\bar d_{R,c}^c l_R)
\ee
where SU(2)$_L$ indices are implicitly contracted with
$\epsilon_{\alpha\beta}$ in the second operator.  The first two 
require
the electric charge of $\Psi$ to be $q_\Psi = 1/6$  while the
last one needs $q_\Psi = 2/3$.
They can be UV-completed
by introducing a color-triplet scalar with couplings
\be
 \epsilon_{abc}\,\epsilon_{AB}(\bar\Psi^c_{A,a}\Psi_{B,b})\Phi_c + 
\Phi^{*a}(\bar u_{R,a}^c l_R,\ \bar L_L^c Q_{L,a},\ \bar d_{R,c}^c l_R)
\label{uvmodel2}
\ee
(we omit writing the dimensionless coupling constants).

$\Phi$ therefore decays like a scalar leptoquark, which can be
consistent with current constraints from LHC if $m_\Phi\gtrsim$ TeV
\cite{Aaboud:2016qeg,CMS:2016qhm,CMS:2016xxl,Romeo:2016abf}.
However, the $\bar\Psi^c\Psi\Phi$ interaction makes it clear that
there is a bound state $\tilde\Phi =\bar\Psi\Psi^c$ with the same quantum numbers
as $\Phi$.  The $\bar\Psi^c\Psi\Phi$ operator becomes an off-diagonal mass
term $\sim f_{\tilde\pi}^2\,\tilde\Phi^*\tilde\Phi$ (where
$f_{\tilde\pi}$ is the hypermeson decay constant) that causes mixing between the elementary
and composite scalars.  Therefore the experimental constraints on
leptoquarks also apply to $\tilde\Phi$, assuming its production cross
section is the same as that of $\Phi$.  

Generally, the production of
$\tilde\Phi^*\tilde\Phi$ will be of the same order as that for
$\bar\Psi\Psi$, depending upon the probability for $\bar\Psi\Psi$
to hadronize into $\tilde\Phi^*\tilde\Phi$ versus other hadron-like
pairs.  In the present case, there are only two ways to hadronize,
either into mesons $\bar\Psi\Psi$ or baryons $\bar\Psi^c\Psi$,
so we expect the production cross section to be about half of that for
an elementary color triplet pair.  Taking this into account, we can
infer the CMS limits on the $\tilde\Phi$ mass to be
$m_{\tilde\Phi}\gtrsim 680\,$GeV if $\tilde\Phi\to \tau b$
predominantly \cite{CMS:2016xxl} and 
$m_{\tilde\Phi}\gtrsim 650\,$GeV $\tilde\Phi\to \mu q$ 
\cite{CMS:2016qhm}.  These are marginally compatible with the
expected value $m_{\tilde\Phi}\sim 750\,$GeV.  The corresponding
ATLAS limits are similar.

\subsection{$N_{\HC}=3$}

If $N_{\HC}=3$ and $\Psi$ carries hypercharge $Y=1$, 
it can couple to right-handed
leptons $e_{R,i}$ of generation $i$, and a neutral colored 
scalar hyperquark $\Phi$,
\be
	\lambda_i\bar\Psi^A_a\, \Phi_A^a\, e_{R,i} + 
	\mu\, \epsilon^{ABC}\epsilon_{abc} \Phi_A^a\, \Phi_B^b\,
	\Phi_C^c
\label{uvmodel}
\ee
Supposing that the scalar is heavy, one can integrate it out
to obtain  a dimension-9 operator schematically of the form
\be
	{\mu\lambda^3\over m_\Phi^6}(\bar\Psi e_R)^3
\label{dim9}
\ee
that allows the charge-3 relic HB to decay into three leptons.
This effective operator was pointed out in ref.\ 
\cite{Craig:2015lra}, where $\Psi$ carried an extra flavor index
$f=1,2$, necessitating the existence of all possible combinations
of $\Psi_1$ and $\Psi_2$ in operators like (\ref{dim9}) to deplete
all the flavors of baryons. We note that the UV-completion
solves another problem of their model, namely the overabundance of
hyperpions of the form $\tilde\pi_{12} = \bar\Psi_1\Psi_2$ that were stable in the 
theory with only the effective operator (\ref{dim9}), but become 
unstable to $\tilde\pi_{12}\to e_R \bar e_R$ by $\Phi$ exchange using
interactions of the type (\ref{uvmodel}).

\subsubsection{Constraints on light $\Phi$}
\label{clp}

If $\Phi$ is relatively light, then analogously to the discussion
in section \ref{lhcconst}, there will be a vector-like 
composite state
$E = \bar\Psi^c\Phi^*$ that has the same quantum numbers as
$e_{R,i}$, and we get mass mixing between $E$ and a linear
combination of the SM leptons $e_{R,i}$. ATLAS has
searched for the decays of a vector-like lepton into $Z$ and a SM
lepton \cite{Aad:2015dha}.  The constraints are not very restrictive,
ruling out the mass ranges 129-176 GeV (114-168 GeV) if the mixing
is primarily to electrons (muons), except for gaps 
144-163 GeV (153-160 GeV) where a heavy lepton is still allowed.

It is also possible to derive constraints from
the rare flavor-violating decays $Z\to \ell^+_i \bar\ell^-_j$.  These
arise because the unitary transformations $U_L$ and $U_R$ that 
diagonalize the $4\times 4$
Dirac mass matrices do not act unitarily in the $3\times 3$ subspace
involving only the SM leptons, which couple to $Z$ whereas
the heavy $E$ state does not. These
flavor-changing couplings are proportional to 
\bea
    {\cal L}_Z = 	-\frac{g}{2c_\W}\bar e_i 
    \Bigg[
	&\!&\!\!\!\!\!\!(U_L^\dagger P_3 U_L)_{ij}(-1+2 s_\W^2)P_L
 \nonumber \\
	&+& (U_R^\dagger P_3 U_R)_{ij}(2 s_\W^2)P_R
	\Bigg] e_j
\label{Zcouplings}
\eea
where $c_\W,s_\W$ are the weak mixing factors and
$P_3$ projects onto the $3\times 3$ subspace of the SM leptons.
Taking $P_3 = 1-P_4$ where $P_4$ projects onto the heavy state,
one can express the nonstandard contributions to (\ref{Zcouplings}) 
as
\bea
\delta{\cal L}_Z = + \frac{g\,f_{\tilde\pi}^2}{2c_\W\, m_E^2}
\bar e_i\slashed{Z}\Bigg[
	&&\!\!\!\!\!(m_\ell\lambda)_i (m_\ell\lambda)_j(-1+2 s_\W^2)P_L
 \nonumber \\
	& +& \lambda_i\lambda_j(2 s_\W^2)P_R \Bigg] e_j
\label{Zcouplings2}
\eea	
where $m_\ell$ is the (diagonal)
light lepton mass matrix and $m_E$ is the composite state mass.
Because of the $m_\ell$-suppresion of the $U_L$ mixing, the 
right-handed couplings dominate.  Assuming that $m_E\sim
f_{\tilde\pi}$, the experimental upper limits
on decays into $e\mu$, $e\tau$, $\mu\tau$ lead to 
$\sqrt{|\lambda_e\lambda_\mu|}\lesssim 0.08$, 
$\sqrt{|\lambda_e\lambda_\tau|},
\sqrt{|\lambda_\mu\lambda_\tau|} \lesssim 0.14$.

Somewhat stronger bounds arise from the diagonal contributions,
which 
can induce flavor nonuniversality in flavor-conserving decays
$Z\to \ell_i \bar\ell_i$ through interference with the SM amplitudes.
Ignoring the small contribution from the left-handed couplings
$g_L$, and assuming that $\lambda_\tau \gg \lambda_\mu, \lambda_e$,
the fractional deviation $\Delta R_{\tau/e} = $BR$(Z\to
\tau\bar\tau)/$BR$(Z\to e\bar e) -1$ is given by 
\be
	\Delta R_{\tau/e} = 2{g_R\,\delta g_R \over g_R^2 + g_L^2}
	= -{8 s_\W^2\over 1- 4 s_\W^2 + 8 s_\W^4}\,
	{\lambda_{\tau}^2\, f_{\tilde\pi}^2\over m_E^2}
\ee
The experimental limit (at $1\sigma$) is $\Delta R_{\tau/e} >
-0.0013$, implying $|\lambda_\tau| \lesssim 0.04$, again assuming that
$m_E\sim f_{\tilde\pi}$.

Products $\lambda_i\lambda_j$ with $i\neq j$ are constrained by
 radiative
flavor violating decays at one loop, illustrated by
 $\mu\to e\gamma$ in fig.\ \ref{mu2egamma}.  Defining
$t=m_\Psi^2/m_\Phi^2$ and writing the transition amplitude 
as $(m_\mu/\Lambda_\mu^2)\bar e_R \sigma^{\mu\nu}q_\nu\epsilon_\mu
\mu_L$, we find \cite{Lavoura:2003xp}
\be
	{1\over\Lambda_\mu^2} = {9e\lambda_e\lambda_\mu\over 32\pi^2 
	m_\Psi^2}
	f(t) < \left(1\over 64\,{\rm TeV}\right)^2
\label{mu2egammaeq}
\ee
where $f(t)$ is as in eq.\ (\ref{b2sgamma}) and the experimental
limit is inferred from refs.\ \cite{Kolda:2009zz,TheMEG:2016wtm}.
Taking for example
$m_\Psi\cong 400\,$GeV  and $f\sim 0.1$ leads to 
the limit $\sqrt{|\lambda_\mu\lambda_e|} < 0.2,$ less restrictive than
that from $Z\to\mu e$.

\subsection{$N_{\HC}=  4$}

For $N_{\HC}= 4$, it is also possible to violate HB with renormalizable
interactions if there exists a colored scalar $\tilde\Phi_{AB,a}$
in the antisymmetric tensor representation of SU($N_{\HC})$, as well 
as a color-triplet fundamental $\Phi^A_a$.  One can then construct the
interactions
\bea
	\epsilon_{ABCD}\,\epsilon_{abc}\,\bar\Psi_{A,a}\Psi^c_{B,b}
	\tilde\Phi_{CD,c} &+& \mu\epsilon_{abc}\tilde\Phi_{CD,a}
	\Phi^C_b\Phi^D_c\nn\\
	 &+& \epsilon_{abc}\Phi^C_a\bar\Psi_{C,b}
	q_{R,c}
\eea
where $q_{R}$ can be either $u_R$ or $d_R$.  Integrating out the
scalars gives the dimension-9 operator
\be
	{\mu\over M_\Phi^4\,M^2_{\tilde\Phi}}\,
	\epsilon_{ABCD}\,\epsilon_{acd}\,\epsilon_{bef}\,
	(\bar\Psi_{A,a}\Psi^c_{B,b})(\bar\Psi_{C,c}q_{R,d})
	(\bar\Psi_{D,e}q_{R,f})
\ee
It allows four $\Psi$'s to decay into two quarks, and would 
mediate decay of the charged hyperbaryon $\Psi^4$, provided
that $\Psi$ has charge $1/3$ or $-1/6$.   The collider phenomenology
stemming from the $\Phi\bar\Psi q_R$ operator 
is similar to that of the model given by eq.\ (\ref{Y23}).

\subsection{Summary}
\label{summary}

We have presented several renormalizable frameworks allowing for
depletion of the unwanted charged relic hyperbaryon.   Most of them
require a charge-neutral scalar hyperquark $\Phi$, that might also be
colored if $N_{\HC}=3$.  If $\Phi$ is sufficiently heavy to avoid being
produced at the LHC, these models are practically unconstrained.  On the
other hand, if $m_\Phi \lesssim$ TeV, $\Phi$ can form a meson-like
bound state with $\Psi$ that has the quantum numbers as a SM quark or
lepton.  We showed that the first case is rather strongly constrained
by ATLAS and CMS searches for heavy quarks (top partners). 

If $N_{\HC}=2$, another possibility is to introduce a scalar 
leptoquark $\Phi$ that is neutral under SU$(N_{\HC})$.  In this model,
there is a bound state of $\Psi\Psi$ with the same quantum numbers
that mixes with $\Phi$ and thus introduces a relatively light
leptoquark state.  This is strongly constrained by ATLAS and CMS,
leaving little room for a model in which the $\bar\Psi\Psi$ hypermeson
could be as light as 750 GeV.

These models typically also predict some level of quark or lepton
flavor violation, but we find that the resulting constraints are
typically weak.  The new Yukawa couplings appearing in 
$\lambda_i\bar\Psi\Phi f_i$,
where $f_i$ is a SM fermion, need only be of order 0.1 in most
cases.  Flavor universality of $Z\to\ell_i\bar\ell_i$ decays
gives the strongest such limit, $\lambda_\tau < 0.04$.

\section{Neutral hyperbaryons}
\label{HBcons}

A second possibility is that the charged HQ can decay into a lighter
neutral HQ, which we denote by $S^A$, that is fundamental under
SU(N$_{\HC}$) and electrically neutral.  In principle it could also carry
QCD color or weak isospin, but as we will show, these options are
generally disfavored by constraints from direct detection of the
resulting hyperbaryon $S^{N_{\HC}}$.  We introduce a scalar $\Phi$
that mediates the decay of $\Psi$ to $S$ plus standard model
particles through an interaction of the form
\be
   \lambda\, \bar S^{A}\,\Phi\,\Psi_{A}
\label{Psidecay}
\ee
followed by decay of the mediator $\Phi$ into standard model
particles.  (Indices corresponding to any additional quantum numbers
are suppresssed here). Alternatively, $\Psi$ and $S$ could be in an
SU(2)$_L$ doublet, so that $\Psi\to S W$ by the SM weak interaction.
In the following, we consider the different possible cases for
additional quantum numbers carried by $S$. 

\subsection{Colored stable hyperquark}
\label{csh}

We first consider the case in which  the neutral HQ $S$ is colored.
The baryonic state that is a singlet under SU($N_{\HC}$) 
is not color-neutral, if $S$ is fermionic.  
The SU($N_{\HC}$) singlet operator
$\epsilon_{A_1,\dots,A_{N}} \bar S^{A_1,a_1}\cdots
\bar S^{A_{N},a_{N}}$ is symmetric under interchange of 
SU(3) indices, and can only be antisymmetric under spin if
$N_{\HC}=2$.  To make a color singlet, it must bind with ordinary quarks,
\be
	B = \epsilon_{A_1,\dots,A_{N}} \bar S^{A_1,a_1}\cdots
       \bar S^{A_{N},a_{N}}\, q_{a_1}\dots q_{a_N}
\label{Beq}
\ee
whose flavors and spins (or spatial configurations) are chosen so as to
make the $q\cdots q$ part of the wave function totally antisymmetric,
while maintaining charge neutrality.  
For example if $N_{\HC}=3$, one can form the antisymmetric 
$s$-wave state of two down quarks and one
up quark, whose flavor/spin wave function is 
\be
	udd \, (\ua\ua\da - \ua\da\ua) + ddu\,(\ua\da\ua - \da\ua\ua)
	+ dud \, (\da\ua\ua-\ua\ua\da)
\ee
This bound state of $SSSudd$ could be expected to behave similarly to
a heavy neutron in its scattering on ordinary baryonic matter.

The scattering properties of dark matter comprised of 
exotic baryonic-like bound states have been discussed in ref.\ 
\cite{Cline:2013zca}.  There it is noted that for low-energy
nucleon-nucleon scattering, the scattering amplitude scales as
${\cal A} \sim 4\pi a/m_N$ where $a$ is the scattering length and
$m_N$ is the nucleon mass.  For scattering in a central potential,
we can expect that $m_N$ represents twice the reduced mass, hence
the amplitude for scattering of a heavy neutron of mass $m_B$ on a 
normal one of mass $m_N$ should scale as 
${\cal A} \sim 2\pi a/\mu$ where $\mu = m_B m_N/(m_B + m_N)\cong
m_N$.  Hence the cross section for $B$-$N$ scattering is roughly four times smaller than that of $N$-$N$ scattering.
(The scattering length is determined by the pion mass and confinement
scale, hence should not depend explicitly upon the mass $m_B$.)

Comparing to the experimentally measured neutron-proton cross section
(see fig.\ 6 of ref.\ \cite{Cline:2013zca}), we can estimate the
cross section for $B$-$N$ scattering at center of mass energy
$\sim m_N v^2\sim 1\,$keV appropriate for direct detection,
namely $\sigma_{BN}\sim 5\,$b.  \jc{This is many orders of magnitude
higher than direct detection limits (spin-dependent or independent), but such strongly
interacting dark matter would be stopped in the earth before reaching
the underground detectors, making such limits inapplicable. 
High-altitude detectors do not suffer from this
limitation \cite{Starkman:1990nj,Mack:2007xj,Erickcek:2007jv}, 
but are too weak to constrain our neutral HB having
only spin-dependent interactions with baryons mediated by pion
exchange.  Instead
one should consider the possibility that these particles will bind to ordinary
matter, creating anomalously heavy isotopes for which stringent
searches have been carried out.  

It is impossible to know whether composite HB's containing ordinary
quarks will bind to ordinary baryons to produce anomalous isotopes,
without doing a nonperturbative calculation such as on the lattice.
However if the HB interacts with nucleons in a similar manner 
as hyperons such as the $\Lambda$ baryon, one could expect the analog 
of the hypertriton, the bound state consisting of $\Lambda$, $p$ and 
$n$, which is known to exist.  Moreover if the HB-baryon interaction is modeled
by pion exchange, then there is always an attractive channel for
fermionic HB's, since the interaction is spin-dependent.  

If HB's do bind to protons, the
abundance of such bound states relative to that of protons is given by
$Y = (m_p/m_{\HB})(\Omega_{\HB}/\Omega_b)$
A search for anomalous hydrogen in sea water finds the
limit $Y < 6\times 10^{-15}$ \cite{Verkerk:1991jf}, leading to the bound
\be
	{\Omega_{\HB}\over\Omega_{\rm\sss DM}}
	\lesssim 10^{-13} \left({m_\HB\over{\rm TeV}}\right) 
\ee
}
In section \ref{relic} we will show that the predicted relic
density is far too large to satisfy this constraint.\footnote{To 
generalize the previous example to other values of $N_{\HC}$, 
one must admit nonzero values  of the HQ electric charge since charge-neutral
combinations of $N_{\HC}$ ordinary quarks do not generally exist.
In this case it may be possible to dispense with $Q$ altogether 
and find a neutral bound state of the form (\ref{Beq}) with $\Psi$
appearing in place of $Q$.  For example with $N_{\HC}=2$, one can 
form the state
\be
	B_\Psi = \epsilon_{AB} \bar \Psi^{A,\alpha}\,
       \bar \Psi^{B,\beta}\, u_{\alpha}u_{\beta}\,	
(\ua\da-\da\ua)
\label{Beq2}
\ee
provided the charges of $\Psi$ and $u$ are equal.  However regardless
of these details, we expect that any bound state containing ordinary quarks
will bind to protons, and the previous result will hold.}

An exception is when $N_{\HC}=3$ with colored scalar HQs, denoted
by $\Phi$.  In that case the bosonic HB state
\be
	B = \epsilon_{ABC}\,\epsilon_{abc}\,
	\Phi^{A,a} \Phi^{B,b} \Phi^{C,c}
\label{scalarB}
\ee 
is neutral under all gauge symmetries and has the correct statistics.
However this model does not have HB conservation as an accidental
symmetry, since the super-renormalizable operator (\ref{scalarB})
can simply appear in the Lagrangian, like in the model of 
eq.\ (\ref{uvmodel}).  We therefore consider it to be unnatural as an
example of HB conservation.

\subsection{Weakly interacting stable hyperquark}
\label{weakrelic}

\begin{figure}[t]
\vspace{0.5cm}
\centering
\centerline{\includegraphics[width=0.5\textwidth]{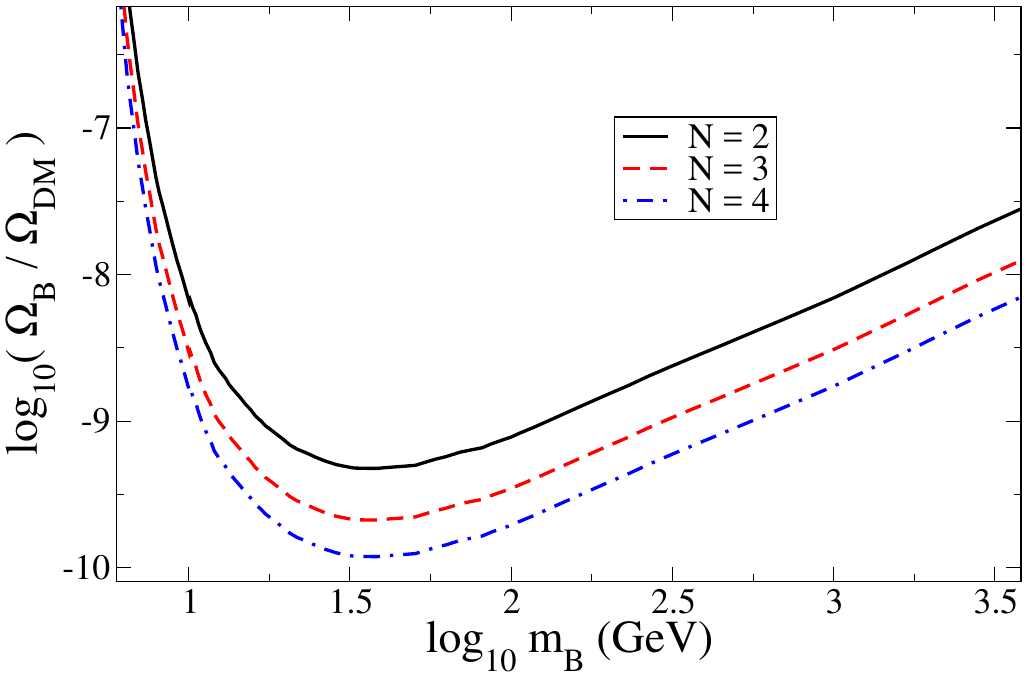}}
\caption{LUX Limit \cite{Akerib:2015rjg,LUX} on fractional density of 
electroweakly-interacting hyperbaryonic dark matter, rescaling
predictions for Dirac neutralino scattering by $Z$ exchange from 
ref.\ \cite{Buckley:2013sca}. }
\label{lux-ew}
\end{figure}

In models where the SM gauge indices of the hyperquarks are embedded
in the fundamental of SU(5) for gauge unification 
\cite{Bai:2016vca,Bai:2016czm,Bai:2015nbs,Harigaya:2015ezk,
Harigaya:2016pnu,Redi:2016kip}, the colored HQ can be expected to decay into a doublet HQ by exchange
of a heavy GUT gauge boson.  The charged component of the doublet can
then decay into the neutral $S$ through weak interactions.  In this
case, the hyperbaryon $S^{N_{\HC}}$ will have weak interactions with ordinary
matter through $Z$ exchange, similar to a hypothetical heavy bound state containing $N$
left-handed neutrinos.   In terms of previously studied models, we
expect the cross section for scattering on nucleons to be similar to
that of Dirac Higgsino dark matter, neglecting the Higgs exchange 
contributions to the scattering that occur in that model and focusing
only on $Z$ exchange.  This has been studied in ref.\
\cite{Buckley:2013sca}, which finds that the spin-independent cross section for
scattering on neutrons is
\be
	\sigma \cong 1.0\times 10^{-37}{\rm\, cm}^2
\ee
This is $\sim (7-9)$ orders of magnitude above the current LUX limit
\cite{Akerib:2015rjg,LUX},
requiring that the relic density of such HBs be correspondingly
depleted.  The cross section for HB scattering is expected to be
$N_{\HC}^2$ times larger due to the number of constituents.  We show
the limit on the fractional abundance as a function of the HB mass
in fig.\ \ref{lux-ew}, including this dependence on $N_{\HC}^2$.

\begin{center}
\begin{table}[b]
\begin{tabular}{|c|c|c|c|c|c|}\hline
$y_\Phi$ & $T_{3,\Phi}$ & $q_\Phi$ &  ${\cal L}_B$ & ${\cal L}_\slashed{B}$ 
& ${{\rm BR}(\tilde\pi\to\gamma\gamma)\over{\rm BR}(\tilde\pi\to gg)}$\\
\hline
$+7/6$ &  $\pm 1/2$ & $2/3,\,5/3$ & 
$\left\{\Phi_\alpha\, \bar Q_L^\alpha\, l_R\atop
\Phi_\alpha\, \bar u_R\, L_L^\alpha\right\}$ & none & 0.12 \\
$+1/6$ & $\pm 1/2$ & $-1/3,\,2/3$ & $\Phi_\alpha\, \bar d_R\,
 L_L^\alpha$ & none & $2.7\times 10^{-3}$\\
\hline
$-1/3$ & 0  & $-1/3$ & $\Phi\, \bar u_R\, l_R^c$ & 
$\left\{{\Phi^*\, \bar Q_L\, Q_L^c,\atop \Phi^*\, \bar u_R\,
d_R^c }\right\}$ & $4.3\times 10^{-4}$ \\
$-4/3$ & 0 & $-4/3$ & $\Phi\, \bar d_R\, l_R^c$ & $\Phi^*\, \bar u_R\, 
u_R^c$ & 0.11\\
\hline
\end{tabular}
\caption{Possible hypercharges, weak isospin and electric charges
of the colored scalar mediator
$\Phi$, the baryon-conserving operators leading to $\Phi$ decay into
SM particles, and allowed baryon-violating operators.  The last column
is the ratio of branching ratios of a pion-like 750 GeV state into
photons versus gluons, for constituents $\Psi$ having the same SM
quantum numbers as $\Phi$.
\label{phidecays}}
\end{table}
\end{center}

\subsection{Singlet stable hyperquark}
\label{singlet}

An interesting possibility is that the scalar mediator carries away
the charge and color of the $\Psi$ hyperquark, leaving $S^A$ charged
only under SU($N_{\HC}$).  Then the possible
couplings of $\Phi$ allowing decays to SM  particles, while
maintaining ordinary baryon number as an accidental symmetry, 
 are limited to the first two cases shown in  table \ref{phidecays}, 
where the hypercharge $y_\Phi$ can take values  $7/6$ or $1/6$.  
The last two, with $y_\Phi=-1/3$ or $-4/3$, are disfavored because
they allow for baryon violation by marginal operators, leading to
rapid proton decay.  
In the favored models, baryon number can be consistently assigned
to all fields, such that $B$ coincides with HB number.

In the preferred models with $y_\Phi=7/6,\,1/6$,  
the scalar $\Phi$ (and
therefore $\Psi$) is an SU(2)$_L$ doublet.  These models are viable
for producing the 750 GeV diphoton signal in the pion-like 
regime, for which the ratio of branching ratios 
$R\equiv$BR$(\tilde\pi\to\gamma\gamma)/$BR$(\tilde\pi\to gg)$ shown in the last column
of table \ref{phidecays} is relevant; this ratio is computed following
ref.\ \cite{Craig:2015lra}. If $R$ is too small, the
observed diphoton rate would require the width for decays into 
gluons to be so large (in order to compensate for the small BR into
photons) that they would exceed the ATLAS bound 
on dijets \cite{Aad:2014aqa}, $\sigma(\tilde\pi\to gg)<2.5\,$pb at
$\sqrt{s}=8\,$TeV.  We find the lower limit $R > 1.6\times 10^{-4}$
to satisfy this constraint; details
are given in appendix \ref{dijet}.    All the models in table 
\ref{phidecays} are consistent with this constraint. 

However these models suffer from another constraint, namely searches
for leptoquarks at the LHC.  Even though $\Phi$ may be very heavy, 
it mixes with bound states of $\tilde\Phi=\bar\Psi S$ that have the same quantum
numbers as $\Phi$.  These hypermesons will be pair-produced at LHC and their
masses must be less than 750 GeV since $m_S < m_\Psi$;
$m_{\tilde\Phi}$ can only
be decreased by mixing with $\Phi$.  They can decay
only into quarks and leptons since they have the quantum numbers of
leptoquarks.  ATLAS and CMS find lower bounds on scalar leptoquarks
decaying into jets and electrons or muons such that $m_{\tilde\Phi} \gtrsim
1\,$TeV for branching ratio $\beta = 100\%$ into one of those channels, and 
$m_{\tilde\Phi} \gtrsim 800\,$GeV for $\beta = 30\%$ 
\cite{Aaboud:2016qeg,CMS:2016qhm,CMS:2016xxl,Romeo:2016abf}.\footnote{CMS
reports a slight excess of $eejj$ events corresponding to 
$m_\Phi=650$, $\beta=0.015$.}
Even if $\Phi$ decays mostly into $\tau$ and 3rd generation quarks,
the limit ranges from $m_{\tilde\Phi} > 500-740\,$GeV for $\beta = 50-100\%$,
from the Run I data.  Thus there is very little parameter space in
which to hide an expected leptoquark with $m_\Phi < 750\,$GeV,
making it difficult to accommodate these models.

\begin{figure}[t]
\vspace{0.5cm}
\centering
\centerline{\includegraphics[width=0.5\textwidth]{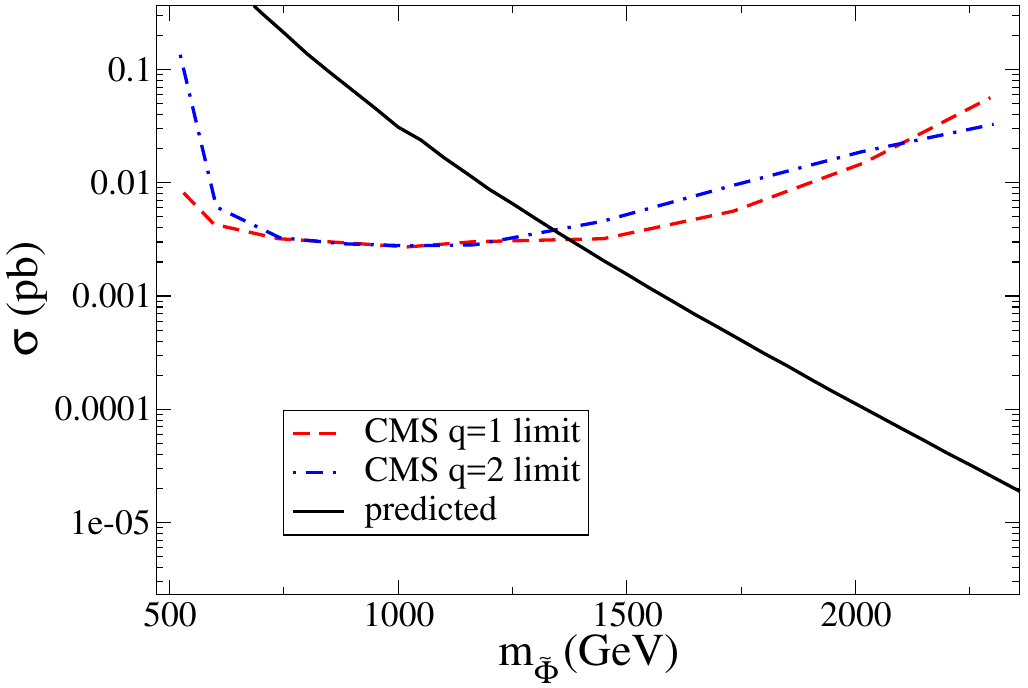}
}
\caption{Predicted LHC production cross section at $\sqrt{s}=13\,$TeV for
for $\bar\Psi\Psi$ pairs that hadronize into leptoquark-like bound
states $\tilde\Phi=\bar\Psi S$, and CMS upper limits for heavy stable particles
of charge 1 and 2.}
\label{lqxsect}
\end{figure}  

A conceivable way out might be to choose small dimensionless couplings 
for (\ref{Psidecay}) and the interactions of table \ref{phidecays} 
such that the composite leptoquark is metastable and decays outside of
the detector (but with a lifetime still below 1 s to avoid problems
with BBN).  CMS has searched for such long-lived charged
particles.\footnote{similar searches by ATLAS are difficult to
interpret in the context of the present model.}\ \  
Ref.\ \cite{Chatrchyan:2012sp} from run I directly
constrains charged hypermesons, which however are assumed to be
produced by Drell-Yan rather than $gg$ fusion.  
Ref.\  \cite{CMS:2015kdx} does a similar analysis in run II, considering DY-produced
particles of charge 1 and 2, obtaining limits
on the production cross section that we reproduce in fig.\
\ref{lqxsect}.  Our prediction for the production of $\tilde\Phi\tilde\Phi^*$ pairs
(assuming they originate from $gg\to\bar\Psi \Psi$ and 
$q\bar q\to\bar\Psi \Psi$ that 
hadronize
mainly into $\tilde\Phi \tilde\Phi^*$) is also shown
there.\footnote{We thank Grace Dupuis for computing this using
MadGraph.}
We expect the leptoquark states of charge $2/3$ and
$5/3$ to be constrained at a similar level to the charge 1 and 2
particles considered in \cite{CMS:2015kdx}, leading to a limit of
$m_{\widetilde\Phi}\gtrsim 1.35\,$TeV, again in contradiction to the
premises of the present model.

\begin{figure}[t]
\vspace{0.5cm}
\centering
\centerline{\includegraphics[width=0.5\textwidth]{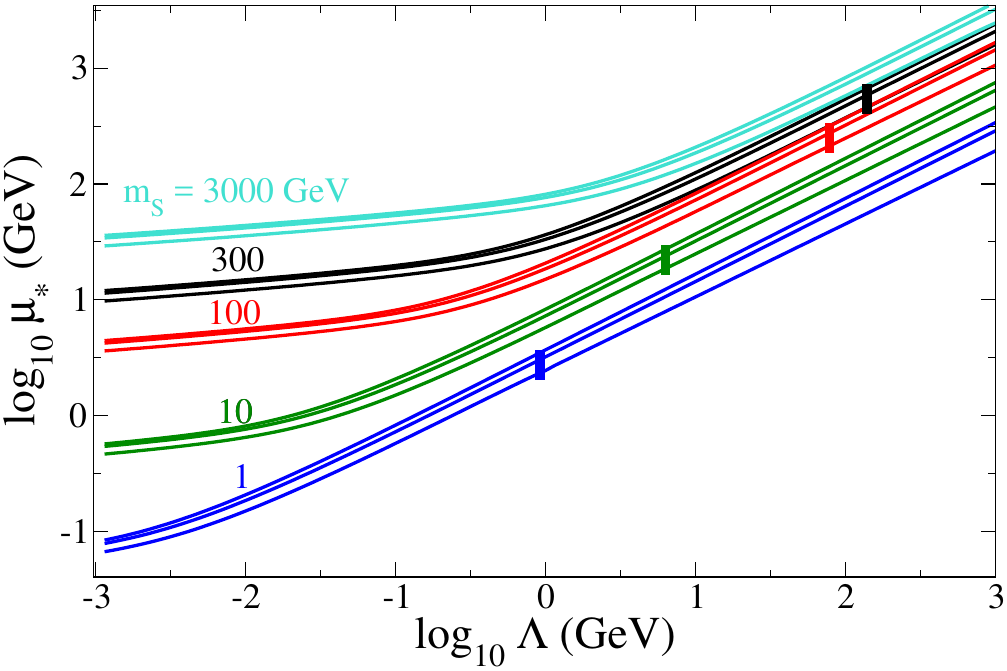}}
\caption{Inverse Bohr radius of the hyperbaryon bound state versus
$\Lambda_\HC$, for several values of the HQ mass $m_S = 3000,\,300,\,100,\,
10,\, 1\,$GeV and $N=2,3,4$ (smaller $N$ gives lower curve for a given
mass.)   Vertical bars show where the perturbative treatment breaks
down and extrapolation of low-$\Lambda$ behavior is used, for a given
$m_S$.}
\label{mustar}
\end{figure}


\section{Relic density of neutral hyperbaryons}
\label{relic}

If hyperbaryon number is conserved and results in a HB that is
neutral under SM gauge interactions, it could be a viable dark matter
candidate.  If it carries weak isospin, then the considerations of 
section \ref{weakrelic} show that it can only be a very subdominant
component of the total dark matter.  In either case, it is interesting
to know what the minimum abundance can be as a result of thermal
freezeout.  Because it has a conserved charge, it is also possible
to have a larger abundance through generation of an asymmetry.  We
leave aside this possibility and consider here the abundance of
the symmetric component, assuming at first that the $S$ hyperquark
has only HC interactions.

\subsection{SU(2)$_L$ singlet hyperbaryons}

The relic HB abundance is sensitive to the ratio $\Lambda_{\HC}/m_S$,
the HC confinement scale over the neutral HQ mass.   If  $m_S
>\Lambda_{\HC}$, there is depletion of the initial $S$ density through
annihilations before confinement, whereas if $m_S < \Lambda_{\HC}$, the
$S$ hyperquarks have a thermal abundance at the confinement phase
transition.    Once confinement begins, a given $S$ has a roughly
equal probability of forming a hypercolor flux string with a
neighboring $S$ or $\bar S$, leading to roughly equal numbers of
hypermesons (that quickly decay away) and HBs.  Following the
confinement phase transition, there can be further depletion of the
HBs by their annihilation.

We estimate the abundance of HBs by solving the Boltzmann equation
in the different regimes of temperature described above.  The
annihilation cross sections for 
$S\bar S\to GG$ (annihilation of HQs into
hypergluons) and of HBs with their antiparticles are needed.
Using refs.\ \cite{Georgi:1978kx,Combridge:1978kx}, we find that the 
first one is
\be
	\langle\sigma v\rangle_{S\bar S\to GG} = {\pi\,\alpha_{\HC}^2(m_S)\over
	4\, m_S^2\, N_\HC^3}(N_{\HC}^2-1)(N_{\HC}^2-2)
\ee
For the gauge coupling we take the four-loop approximation of
ref.\ \cite{Chetyrkin:1997un} with $n_f=0$ flavors, since we are
interested in running only up to the scale $m_S$, presumed to be the
lightest HQ mass in the theory. 

For the annihilation of HBs, it is difficult to estimate the 
cross section due to the strong dynamics and the fact that the
HBs are composite states.  One possibility is to 
use the geometric cross section
\be
	\langle\sigma v\rangle_{\rm geo} = {\pi\over\mu_*^2}  
\ee
where $\mu_*$ is the inverse Bohr radius of the HB, estimated along
the lines of ref.\ \cite{Alves:2010dd}.\footnote{We disagree with
their
numerical coefficient for the expectation value of $1/r_{ij}$ for the
Coulomb-like contribution to the potential.}\ \   
To extend their method to
the regime of strong coupling, we add a linear confining potential
$\sum_{i<j}c\,\Lambda_{\HC}^2 r_{ij}$ to the Coulomb-like term, to 
obtain the expectation
value of the Hamiltonian for a hydrogen-like wave function $\psi\sim
e^{-\mu_* r/2}$ as
\bea
	\frac{\langle H\rangle}{N_{\HC} }&=& \frac{\mu_*^2}{8 m_S} - 
	\frac{5(N_{\HC}-N_{\HC}^{-1})}{64}\alpha_{\HC}(\mu_*)\mu_* 
	\nonumber\\
	&+& \frac{35\,c\,(N_\HC-1)\Lambda_\HC^2}{16\mu_*}
\label{Havg}
\eea
where the value $c=1.9$ is inferred from refs.\ 
\cite{Guagnelli:1998ud,Gockeler:2005rv,Teper:1998kw} for the
case of $N_{\HC}=3$.
Minimizing (\ref{Havg}) with respect to $\mu_*$ gives an implicit 
equation for $\mu_*$ that can be solved by iteration.  The resulting
$\mu_*$ as a function of $\Lambda_\HC$ is shown for several values of
$m_S$ in fig.\ \ref{mustar}.
This procedure
breaks down when $\Lambda \gtrsim m_S$ because the gauge coupling
becomes nonperturbative and the middle term in (\ref{Havg}) diverges
to large negative values as $\mu^*\to \Lambda_\HC$ from above.  The approximate values of $\Lambda_\HC$
where this starts to occur are indicated by heavy dots
in fig.\ \ref{mustar}.  Since the dependence of $\log_{10}\mu_*$
on $\log_{10}\Lambda_\HC$ is very linear (corresponding to the power
law $\mu_*\sim \Lambda_\HC^{0.63}$) in the regions below the
dots, we use linear extrapolation to extend our predictions to 
higher values of $\Lambda_\HC$.   The available
final states for HB annihilation almost always include the hypermesons
$\tilde\pi\tilde\pi$, even when they are more quarkonium-like than pion-like
(the regime of $m_S\gg\Lambda_{\HC}$).  The only exception is when $N=2$ so
that mesons and baryons have the same number of constituents.

To compute the relic HB abundance, we numerically solve the Boltzmann
equation starting from high temperatures using the $S\bar S\to GG$
cross section, evolving down to the confinement temperature $T=\Lambda_\HC$.
We assume the confinement transition occurs rapidly and that the
initial abundance of HBs at $T=\Lambda_\HC$ is related to that of
HQs by $Y_B = Y_Q/(2 N_\HC)$.  Taking this as the initial condition
for the Boltzmann equation using the HB annihilation cross section,
we evolve $Y_B$ to its freeze-out temperature.  This approach is
generally necessary, rather than the usual analytic approximations,
because of the unusual thermal history that often occurs: at the
confinement transition, HBs can often be produced starting with a
density far exceeding the equilibrium abundance (since the HB mass
is $\sim N_\HC$ times $m_S$).  Then the HB annihilations can be in
equilibrium, even though a naive treatment would imply that they
froze out already at an earlier temperature.

\begin{figure}[t]
\vspace{0.5cm}
\centering
\centerline{\includegraphics[width=0.5\textwidth]{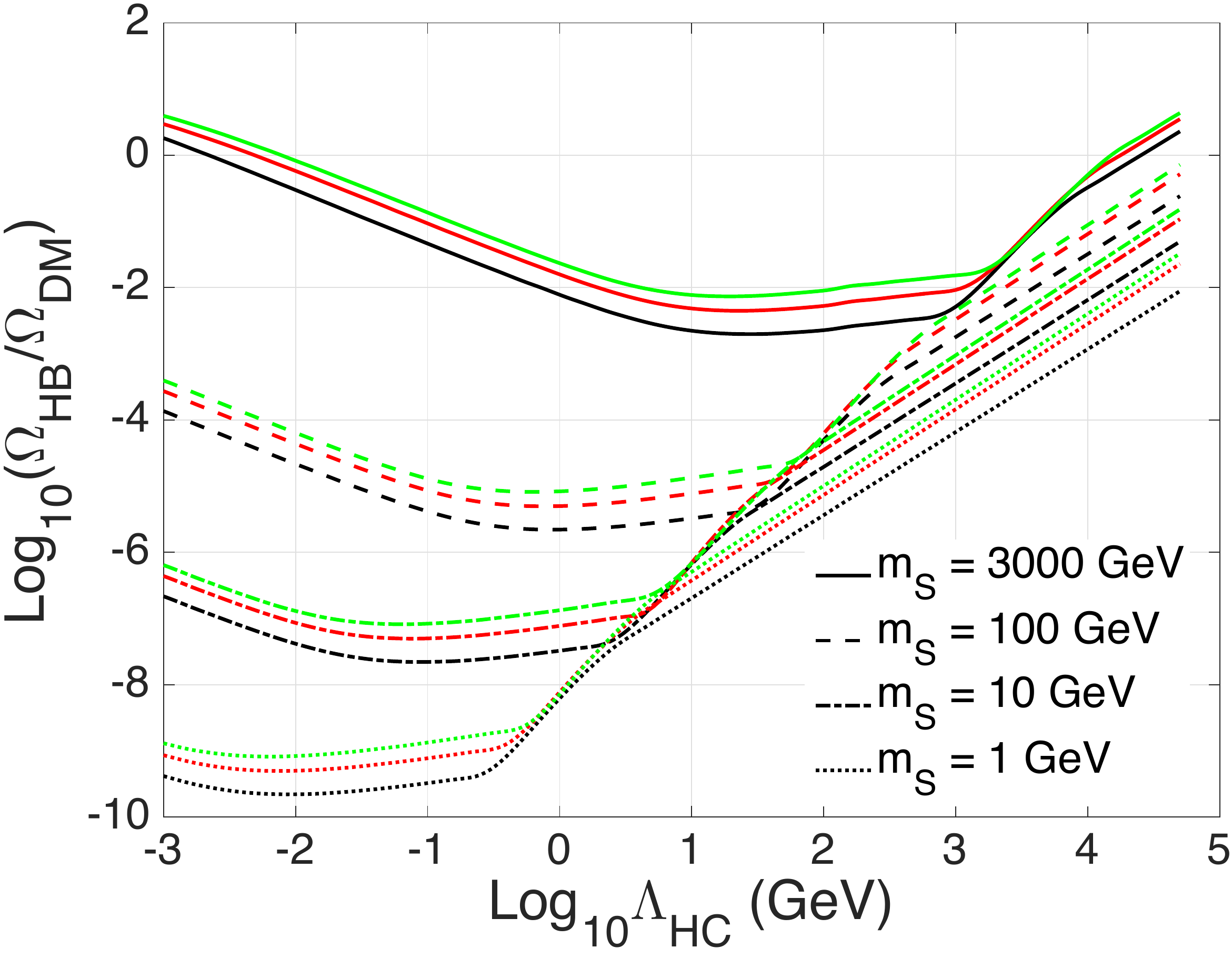}}
\caption{Estimated relic abundance of the lightest neutral 
hyperbaryon versus confinement scale $\Lambda_\HC$, for 
neutral HQ masses $m_S=1,\,10,\,100,\,3000\,$GeV (from bottom to top), 
and $N_{\HC}=2,3,4$ (from bottom to top for
each $m_S$).  
}
\label{relic}
\end{figure}

The results are shown in fig.\ \ref{relic}, where the fractional
abundance of HBs relative to the total observed dark matter is plotted
as a function of $\Lambda_\HC$ for several HQ masses and
$N_\HC=2,3,4$.  From conventional thermal freezeout one expects that
 $\Omega_\HB\sim 1/(\sigma
v)$ as a function of $\Lambda_\HC$.  This explains the simple
power-law behavior of $\Omega_\HB$ in fig.\ \ref{relic}(a), since
$\mu^*\sim\Lambda_\HC^{0.63}$ at large $\Lambda_\HC$.  
 At small
$\Lambda_\HC$ the trend is different because at the confinement
temperature $T_c= \Lambda_\HC$, the initial HB abundance is 
much higher than the equilibrium abundance.  In this situation
the conventional dependence is not applicable and we find the 
different behavior $\Omega_\HB \sim 1/(\Lambda_\HC\,\sigma
v)$  in fig.\ \ref{relic}.

We find that unless $m_S$ is near the TeV scale (hence not relevant
for a diphoton signal at 750 GeV), the symmetric HB component can
provide only a subdominant contribution to the total dark matter
density.  Although the observed density is obtained at low 
$\Lambda_\HC\sim 3\,$MeV, this region is not viable because of
the presence of very light and long-lived glueballs that will
disrupt big bang nucleosynthesis for $\Lambda_\HC\lesssim$ several
GeV\cite{Iwamoto:2016ral}.  But at large $\Lambda_\HC$,
it is possible to obtain the observed abundance.
For $m_S\cong 3\,$TeV and $\Lambda_\HC\cong 20\,$TeV for
example, we find that HBs can constitute all the dark matter, even
with no asymmetry.  We expect the geometric cross section to provide
a lower limit on the true annihilation cross section, which might
require a dedicated lattice study to determine with greater
certainty.  Hence the actual abundance might be smaller than our 
estimate, although we expect the qualitative dependences to
hold.  Even   a highly subdominant component of
HB dark matter could still lead to observable consequences if the HQs
have standard model weak interactions, as we describe next.

\subsection{SU(2)$_L$ doublet hyperquarks}

As mentioned in section \ref{weakrelic},  one way in which the charged
HQ $\Psi$ could decay into the neutral one $S$ is  by embedding $\Psi$
and $S$ into a fundamental of $SU(5)$ for GUT. This leads to strong
constraints on the relic abundance from direct detection.  But the
same weak interactions could in principle have an impact on the 
abundance through the annihilations into SM particles.  The possible
two-body final states include $ZZ, WW, Zh$ and $ f\bar{f}$, depending
on the mass of the HQs or HBs. More details on annihilation cross
sections are given in appendix~\ref{Anni_Xsec-app}.

However we find that these extra annihilation channels have a'
negligible effect on the HB abundance, being much weaker than
the hypercolor interactions; we can therefore infer the densities
from fig.\ \ref{relic}.  Comparing to the direct detection constraints
shown in fig.\ \ref{lux-ew}, it can be seen that very light HBs 
made from HQs with mass $m_S\sim 1\,$GeV$\gtrsim\Lambda_\HC$ can
be compatible with the constraints, but heavier ones are ruled out
by several orders of magnitude.

\begin{figure}[t]
\vspace{0.5cm}
\centering
\includegraphics[width=0.5\textwidth]{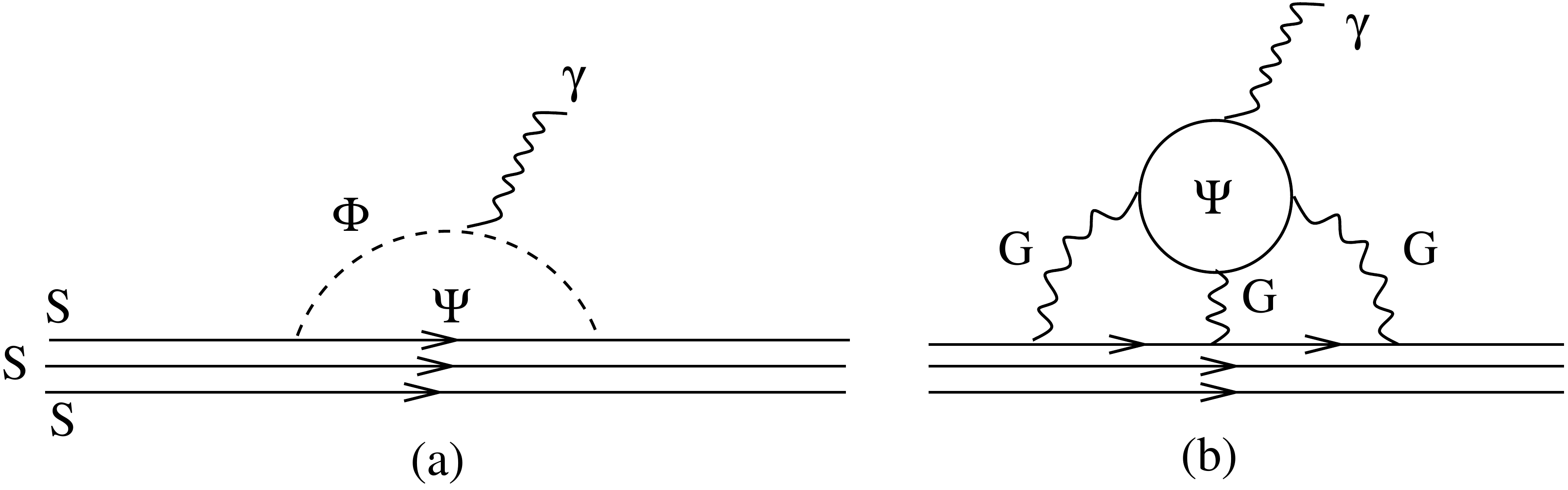}
\caption{Diagrams that induce a magnetic dipole for the neutral hyperbaryon.
$G$ denotes the hypergluon in (b).}
\label{dipole}
\end{figure}

\begin{figure}[t]
\vspace{0.5cm}
\centering
\includegraphics[width=0.5\textwidth]{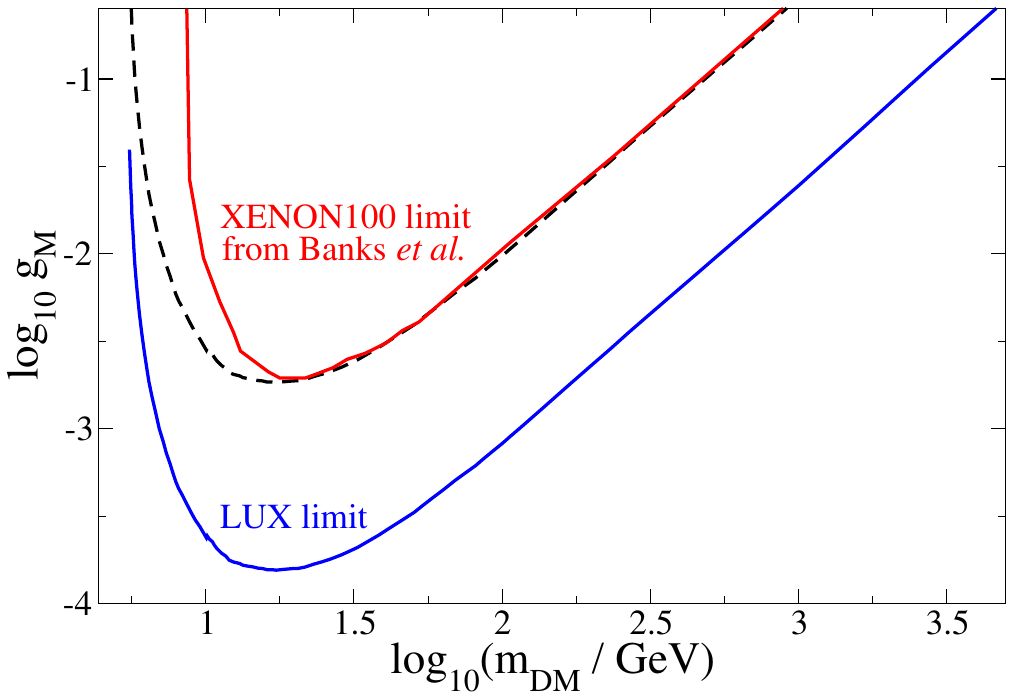}
\caption{Upper limit from direct detection on gyromagnetic
ratio for magnetic dipolar dark matter, found by updating results
of ref.\ \cite{Banks:2010eh}.}
\label{glimit}
\end{figure}

\section{Electromagnetic interactions of neutral relics}
\label{EM}
Even if the neutral hyperbaryons have no residual strong or
weak interactions
with nuclei, as would states of the form (\ref{Beq}), they inevitably
have electromagnetic interactions through the diagrams shown in 
fig.\ \ref{dipole}.  Although this turns out to be unimportant for
direct detection, for
completeness we discuss their effects.  
We estimate these diagrams as giving
magnetic dipole moments for $S$ of order
\be
	\mu_S = {e\,\lambda^2\, f\over 16\pi^2\, m_\Psi},\quad
	 {e\,\alpha_{\HC}^3(m_\Psi)\over 1152\pi^3\, m_\Psi}
\label{muB}
\ee
where $f$ is a function of mass ratios, of order $-0.2$ for models
of interest here.  Details
are given in appendix \ref{dipole-app}.  The magnetic moment of the
hyperbaryon is then $\mu_B \cong N_{\HC}\mu_S$.  The above estimates
assume that $\Psi$ is the heaviest particle in the loop.  In the
pion-like regime where $m_\Psi \ll \Lambda$, the quark model shows
that $m_\Psi$ should be interpreted as the constituent quark mass
rather than the current quark mass.  For our estimates we thus take
it to be $\sim 375\,$GeV.

The limit on the magnetic dipole moment from direct detection can be
parameterized by writing it in terms of the gyromagnetic ratio
$g_M$,
\be
	\mu_B = {g_M e \over 4 m_B}
\ee
The limit on $g_M$ was found in ref.\ \cite{Banks:2010eh} using
data of XENON100 \cite{Aprile:2010um}.  Updating this limit using the
current LUX bounds \cite{Akerib:2015rjg,LUX}, we find the result shown 
in fig.\ \ref{glimit}.  To rescale the limit from XENON100 to LUX,
we notice that the sensitivity of both experiments scales with
dark matter mass in the same way for $m_{DM} > 20\,$GeV, while LUX has
greater relative sensitivity at lower masses, as shown by the
dashed curve in fig.\ \ref{glimit}.  In the large mass
region, the LUX limit on the spin-independent scattering cross section
is 140 times lower than that of XENON100, so we rescale the limit on
$g_M$ by a factor of $1/\sqrt{140}$, taking into account the greater
sensitivity of LUX at lower masses.  

The limit on $g_M$ assumes that the dark matter candidate has the
full relic density, which as we have seen in section \ref{relic}
need not be the case.  For the three-loop contribution, we can 
use this to constrain $\Omega_{HB}/\Omega_{DM}$ as a function of 
$\Lambda$ and $m_S$ since all the quantities entering into $\mu_S$
are determined.  However the resulting upper limit is always
greater than unity, so this provides no meaningful constraint.
For the one-loop contribution, we can insert the minimum fractions
of the total DM density found in section \ref{relic} to get an upper
bound on the coupling $\lambda$ that induces decay of $\Psi$ to
$S$ plus SM particles.  Again, the upper limits on $\lambda$ are hardly
constraining, being greater than unity in all cases.

\section{Uncolored models}
\label{nocolor}

Although not favored by the compatibility of $\sqrt{s}=8\,$TeV
versus 13 TeV LHC data, a number of authors have shown that
the diphoton signal in the 13 TeV data can be accommodated by purely 
electromagnetic
production through photon fusion.  The charged hyperquark $\Psi$ is
then neutral under SU(3)$_c$.  If HB number is conserved, then it is
compulsory to have neutral hypercolored particles to prevent charged
stable hyperbaryons; so we will assume the model is extended with a
neutral spinor $S$ and a scalar $\Phi$ as previously.  Then
$\Psi$ can decay by via $\Psi\to S f_i\bar f_j$ through interactions
of the form
\be
	\lambda\, \bar S^A\,\Phi\,\Psi_A  + \lambda_{ij} \bar f_i \Phi
	f_j
\label{med_int}
\ee
where $f_{i,j}$ are standard model fermions. There are two ways
of choosing combinations of SM fermions consistent with gauge
invariance, listed in table \ref{tab2}.  They correspond to
electric charges of
$1$ or $2$ for the mediator and hence of the $\Psi$, and they imply
that $\Phi$ carries lepton number 2.  We can consistently assign
the same lepton number to $S$ so that overall lepton number is 
conserved by the new interactions.  Depending upon the generational
structure of the couplings $\lambda_{ij}$ however, there could be
violations of individual flavor conservation, or of 
lepton flavor universality.  For example the considerations of 
section \ref{clp} give 
$\sqrt{|\lambda_{\mu\mu}\lambda_{\mu e}|} < 0.6,$ for the same choices of
mass spectrum, weaker by the factor of $(3N_\HC)^{1/2}$ for the 
lack of color/hypercolor in the loop.

These models are similar to the favored ones discussed in section
\ref{singlet} in terms of constraints on the relic HB; they are
cosmologically safe, with subdominant relic densities as predicted by
figure \ref{relic} and unimportant interactions with normal matter
through their small loop-induced magnetic moments.  Since the scalar
mediator couples to lepton pairs instead of being a leptoquark,  
pairs of charged mesons $\bar\Psi S$ decaying to monoleptons (in the
case of the $\bar L L^c$ coupling where one of the particles is a
neutrino) or same-sign dileptons would be a collider signature at
partonic center of mass energies below 1.5 TeV.  

The single-lepton signal is constrained by ATLAS and CMS searches for
events with one lepton and missing transverse energy
\cite{Aaboud:2016zkn,Khachatryan:2014tva}. The more recent ATLAS
result limits $m_S\gtrsim 4\,$TeV if the couplings $\lambda_{ei}$ or
$\lambda_{\mu i}$ are of the same order as the SU(2)$L$ gauge
coupling.  The dilepton channel is relatively unconstrained, since
\mbox{ATLAS} and CMS searches for same-sign dileptons have so far also
required the presence of jets.

\begin{center}
\begin{table}[t]
\begin{tabular}{|c|c|c|}\hline
$\bar f_i f_j$ & $\epsilon_{ab}\bar L_a L^c_b$ & $\bar l_R l^c_R$ 
\\
\hline
$q$ & 1 & 2  \\
\hline
\end{tabular}
\caption{Possible combinations of SM fermions coupling to the 
bosonic mediator $\Phi$, and their charges, for models with
uncolored hyperquarks.
$L$ and $l$ stand for SU(2)$_L$ doublet and singlet leptons,
respectively.
\label{tab2}}
\end{table}
\end{center}

\section{Conclusions}
\label{conclusions}

An additional SU($N_{\HC}$) gauge group factor is a plausible
and economical extension of the standard model.  If new matter fields
transform in the fundamental of SU($N_{\HC}$), the analog of baryon
number in the new sector is an issue: if it is conserved then the
properties of relic particles must be considered, while if it is
broken through renormalizable interactions, interesting constraints
can arise from LHC searches for decays associated with these new 
interactions.  

Our considerations are most relevant for models similar to those  that
can explain  the tentative LHC diphoton excess, where we assumed that
a charged hyperquark in the fundamental of SU($N_{\HC}$) also carries
QCD color.  Such a particle must decay into standard model states and
possibly a neutral state that allows for hyperbaryon number to be
conserved, and which is a dark matter candidate.  In the case that HB
is not conserved, we identified a limited range of renormalizable
models, that are summarized in  section  \ref{summary}.  Even if the
750 GeV diphoton excess at LHC is only a statistical fluctuation,
models consistent with it provide a benchmark for what
could be close to the current sensitivity of ATLAS and CMS.

If HB is
conserved, the renormalizable models  consistent with direct detection
constraints and normal baryon  conservation are also limited, and turn
out to be significantly constrained by LHC leptoquark searches.  The
viable models have a charged hyperquark $\Psi$ and a scalar mediator
$\Phi$ with quantum numbers $(3,2,7/6)$ or $(3,2,1/6)$ under
SU(3)$_c\times SU$(2)$_L\times$U(1)$_y$, while the neutral hyperquark
$S$ is a singlet.  An interesting feature of these models is the
presence of composite $\bar\Psi S$ leptoquarks (in addition to the
heavy fundamental scalar leptoquark $\Phi$), that must have masses $\gtrsim$ 1
TeV to satisfy LHC constraints.

An aspect of our work that transcends diphoton signals is the more
general possibility that dark matter is a baryon-like state of a new
confining sector.  The relic density computation for the symmetric
component is complicated by the first-order confinement phase
transition of the SU($N_{\HC}$) sector.   We find that hyperquark
masses and confinement scales below a TeV, the density is generally
much smaller than the observed dark matter density, but for
$m_S\sim 3\,$TeV, $\Lambda_\HC\sim 20\,$TeV, it could account for
all of the dark matter.
\jc{Searches for anomalous isotopes strongly disfavor the
$S$ hyperquark from being colored under
QCD, and if it is part of an
SU(2)$_L$ doublet, direct dark matter searches} limit its abundance to be $\lesssim 10^{-8}$
of the observed DM density, depending upon the hyperbaryon mass.
Otherwise the interactions of hyperbaryons with nuclei arise only
through loops and give very weak constraints on the model parameters.
Such models would be probed more directly through the collider
constraints as discussed above.

{\bf Note added:} as we were completing this work, ref.\ 
\cite{Hertzberg:2016jie} appeared, which treats astrophysical
constraints on models similar to those we have considered, but in the
case where the charged HQ $\Psi$ is stable and binds with ordinary
quarks to make a neutral relic.  According to our analysis in section
\ref{csh}, \jc{such relics are disfavored by anomalous isotope 
searches,}
which were not considered in ref.\ \cite{Hertzberg:2016jie}.

\bigskip
{\bf Acknowledgments.}  
This work was performed in part at the Aspen Center for Physics, which is supported by National Science
Foundation grant PHY-1066293.
We thank B.\ Cyr for collaboration at an early stage, 
J.\ Cornell, C.\ Kilic and J.\ Berger for helpful discussions,
and G.\ Dupuis for assistance with MadGraph.  
WCH is supported by the Postgraduate Scholarship Program of 
China Scholarship Council (No.201504910563).  JC acknowledges support
from NSERC (Natural Sciences and Engineering Research Council of
Canada) and FRQNT (Fonds de recherche du Qu\'ebec -- Nature et
technologies).  The research of GDM is supported by Land Hessen.

\appendix
\section{Dijet constraint}
\label{dijet}
Here we derive a lower bound on the ratio of branching ratios
$R = {\rm BR}(\tilde\pi\to\gamma\gamma)/{\rm BR}(\tilde\pi\to
gg)$ from the observed LHC diphoton excess and the upper limit on
dijet production.  The total cross section for $pp\to\tilde\pi$
by gluon fusion is \cite{Franceschini:2015kwy}
\be
	\sigma(pp\to\tilde\pi) = {1\over s}{\Gamma_{\tilde\pi}
	\over m_{\tilde\pi}}
	\times\left\{174, \sqrt{s}=8{\rm\ TeV}\atop 2137, 
	\sqrt{s}=13{\rm\ TeV}\right.
\ee
while the cross section for $pp\to{\tilde\pi}\to\gamma\gamma$ at 
13 TeV is
\be
	\sigma(pp\to{\tilde\pi}\to\gamma\gamma)= 5\times 10^6{\rm\, fb}\cdot
	{\Gamma({\tilde\pi}\to\gamma\gamma)\over m_{\tilde\pi}} \cong 5{\rm\, fb}
\label{gamgam13TeV}
\ee
to match the central experimental value.  Similarly, the
cross section for $pp\to{\tilde\pi}\to gg$ at 
8 TeV is
\be
\sigma(pp\to{\tilde\pi}\to gg)= 4\times 10^5{\rm\, fb}\cdot
{\Gamma({\tilde\pi}\to gg)\over m_{\tilde\pi}} < 2.5{\rm\, pb}
\label{gg8TeV}
\ee
taking account of the 174/2137 reduction in gluon luminosity at
8 TeV, and quoting the experimental dijet limit of 2.5 pb 
\cite{Aad:2014aqa}.  Taking the ratio of (\ref{gamgam13TeV}) and
(\ref{gg8TeV}) gives the lower limit $R > 1.6\times 10^{-4}$.

\begin{figure}[t]
\vspace{0.5cm}
\centering
\includegraphics[width=0.5\textwidth]{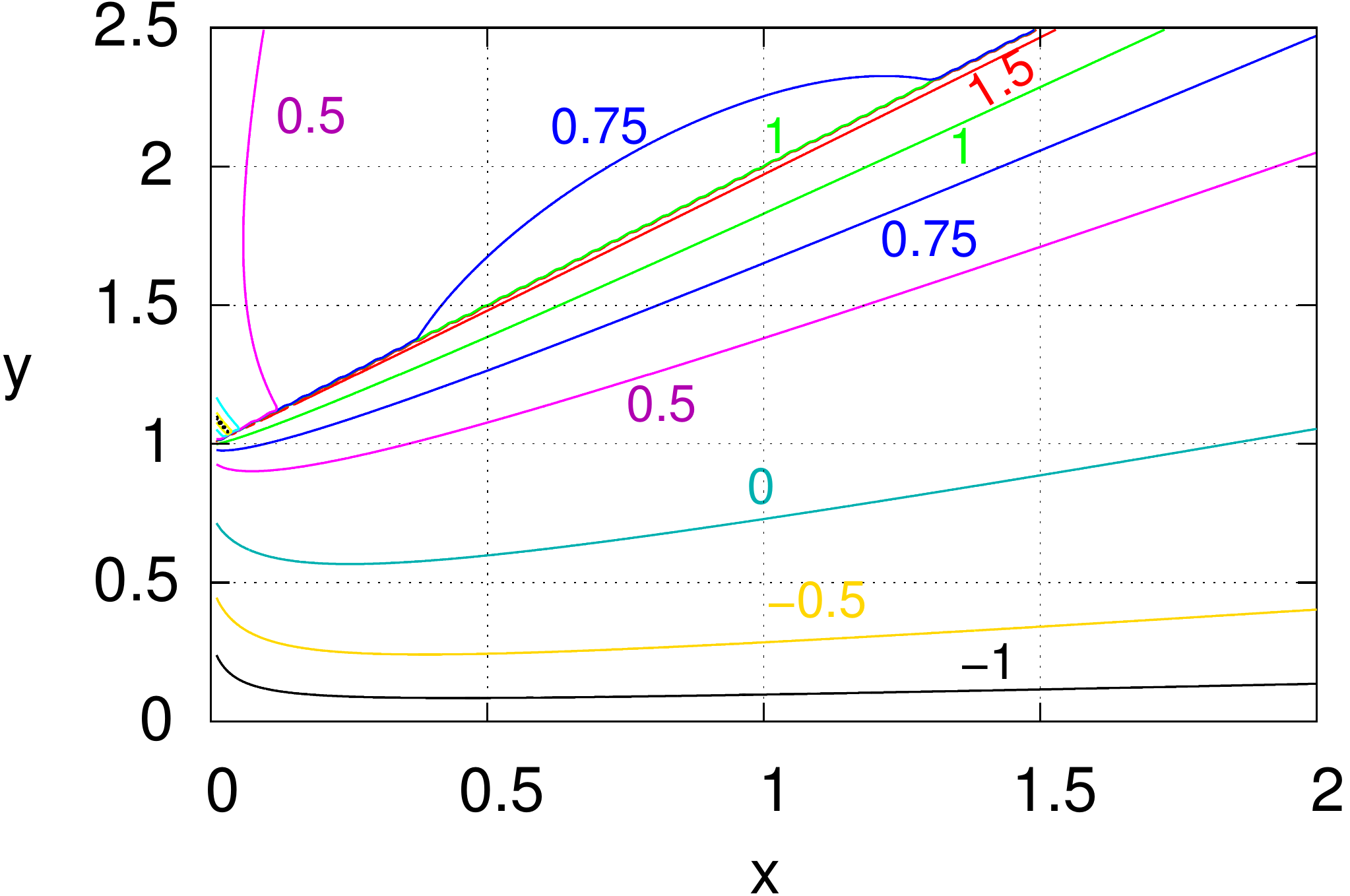}
\caption{Contours of $\log_{10}|f|$ for the
loop function $f$ from eq.\ (\ref{ffun})
for the dipole moment of neutral hypequarks.}
\label{dipolef}
\end{figure}

\section{Dipole moments}
\label{dipole-app}

In this appendix we estimate the loop-induced interactions of neutral
hyperbaryons with photons, relevant for direct detection.  The
interactions are shown in fig.\ \ref{dipole}.  The one-loop diagram
can be computed exactly, giving
\be
\mu_B = {q_{\Phi} e |\lambda|^2 f \over 32\pi^2 m_{\Psi}}
\ee
where $f$ is the loop function,
\bea
f &=& 2 + \frac{1 - x (x + y)}{y^2} 
\log x^2 \nonumber\\
&-& \frac{2[(x + y)^2 - 1](1 - x^2 + xy)}{y^2 
r} \ln\frac{2 x}{ x^2 - y^2 + 1 - r
},\nonumber\\
r &=& \sqrt{[(x - y)^2 - 1][(x + y)^2 -1]}
\label{ffun}
\eea
with $q_\Phi$ the electric charge of $\Phi$, $x = m_\Psi / m_\Phi$
and, $y = m_S / m_\Phi$.
We plot $f$ in fig.\ \ref{dipolef}.

To estimate the
three-loop diagram, we start by integrating out the $\Psi$
hyperquark to obtain an Euler-Heisenberg-like effective Lagrangian 
for the photon-hypergluons vertex \cite{Groote:2001vr} 
\be
\mathcal{L} =  {e\, g_H^3\, d_{abc} \over180\, m_\Psi^4 (4\pi)^2} 
\left[ 14 \mathrm{tr} (F G^a G^b G^c) - 5 \mathrm{tr}(F G^a) \mathrm{tr}(G^b G^c) \right]
\ee
where $F_{\mu\nu}, G^{a}_{\mu\nu}$ are field strength tensors for 
photon and hypergluon respectively, $d_{abc} =  2 \mathrm{tr 
(\{t_a, t_b\}t_c)}$ is the totally symmetric structure constant 
for SU$(N_{\HC})$, and the traces are taken with respect to the 
Lorentz indices.  Then the color factor for the dipole moment 
diagram is ${ d_{abc}\,\mathrm{tr} (t^a t^b t^c) / N_{\HC}} = 10/9$ 
if $N_{\HC} = 3$.  We roughly estimate the effect of the Lorentz
structure and the additional two loops as giving a factor of
$9/(16\pi^2)^2$ to the magnetic moment, 
\be
	\mu_B \sim {e\, g_H^6\over 180\, m_\Psi^4 (4\pi)^2} \cdot{10\over 9}\cdot{9\over(16\pi^2)^2}
	= {e\,\alpha_H^3\over 1152\pi^3\, m_\Psi}
\ee
Since the limits from direct detection on this operator are very weak,
it is unlikely that a more accurate computation would change our 
conclusions.

\section{Annihilation cross section for doublet $S$}
\label{Anni_Xsec-app}
If the neutral hyperquark $S$ is in an
SU$(2)_L$ doublet, it has additional channels for annihilation into
SM states $ZZ, WW, Zh$ and $f\bar f$, with purely
vectorial couplings to gauge bosons due to the
vector-like nature we assumed here. 
The relevant $s$-wave cross sections
are
\bea
\langle\sigma v\rangle_{\sss ZZ} &=& 
{g^4\over 64\pi c_\W^4 m_S^2} {(1 - x_\Z^2)^{3\over2} 
\over (2 - x_\Z^2)^2} \label{ZZann}
\\
\langle\sigma v\rangle_{\sss WW} &=& {g^4 \over 64\pi m_S^2} {(1- x_\W^2)^{3\over2} } \Big[{4(4 + 20 x_\W^2 + 3 x_\W^4) \over (4 - x_\Z^2)^2 }   \nonumber \\
&-& {4 (4x_{\sss \Psi} + 10 x_\W^2(1+x_{\sss \Psi}) + 3 x_\W^4)  \over (1 + x_{\sss \Psi}^2 - x_\W^2) (4 - x_\Z^2)} \nonumber \\
&+& {4x_\W^2(x_{\sss \Psi}^2 + 3x_{\sss \Psi} + 1) + 4 x_{\sss \Psi}^2 + 5 x_\W^4 \over (1 + x_{\sss \Psi}^2 - x_\W^2)^2} \Big]	\label{WWann} \\
\langle\sigma v\rangle_{\sss Zh} &=& {g^4 [(4 - x_\h^2)^2 + 2 x_\Z^2(20 - x_\h^2) + x_\Z^4] \over 4096\pi c_\W^4 m_S^2 (4 - x_\Z^2)^2}  \nonumber \\
&\times& \sqrt{(4-x_\Z^2)^2 - 2 x_\h^2(4 + x_\Z^2) + x_\h^4}
\label{Zhann} \\
\langle\sigma v\rangle_{\sss ff} &=& {g^4\over 8\pi c_\W^4 m_S^2} {(1 - x_\f^2)^{1\over2} \over (4 - x_\Z^2)^2} \nonumber \\
&\times& [g_{Vf}^{2} (2 + x_\f^2) + 2 g_{Af}^{2}(1 - x_\f^2)] \label{ffann}
\eea
where $c_\W \equiv \cos\theta_W, x_i = {m_i/m_S} (i=Z,W,h,f,\Psi)$ and $g_{Vf}\,(g_{Af})$ is the (axial) vector coupling of the fermion $f$ to $Z$ boson. For hyperquarks in fundamental of SU$(N_{HC})$, there is an extra factor of $1/N_{HC}$ for the above formulas taking account of averaging initial degrees of freedom.

For a hyperbaryon with $N_{HC}$ hyperquarks, we expect that there is a
coherent enhancement factor of $N_{HC}$ for each  gauge interaction
vertex of the HB. Then the $s$-wave annihilation cross section for HB
can be rescaled from eq.~(\ref{ZZann}-\ref{ffann}) as
\bea
\langle\sigma v\rangle_{\sss ZZ}^B &=& {4\over(N_{HC} +1)^2} N_{HC}^4\langle\sigma v\rangle_{\sss ZZ}\\
\langle\sigma v\rangle_{\sss Zh}^B &=& {4\over(N_{HC} +1)^2} N_{HC}^2\langle\sigma v\rangle_{\sss Zh}\\
\langle\sigma v\rangle_{\sss ff}^B &=& {4\over(N_{HC} +1)^2} N_{HC}^2\langle\sigma v\rangle_{\sss ff}
\eea
with $m_S$ replaced by $m_B$ and $x_i$ by $y_i = m_i/m_B (i=Z,W,h,f,B_\Psi)$. The factor of
${4/(N_{HC} +1)^2}$ corrects for the averaging over spin degrees of 
freedom of the HB.
For the $WW$ final state, the rescaling is more complicated 
because of interference between the $s$- and $t$-channel 
annihilations,
\begin{widetext}
\bea
\langle\sigma v\rangle_{WW}^B &=& {4 N_{HC}^4\over(N_{HC} +1)^2} {g^4\over 64\pi m_B^2}{(1- y_\W^2)^{3\over2} } 
\Big[ {4y_\W^2(y_{\sss \Psi}^2 + 3y_{\sss \Psi} + 1) + 4 y_{\sss \Psi}^2 + 5 y_\W^4\over (1 + y_{\sss \Psi}^2 - y_\W^2)^2}
 - {4 (4y_{\sss \Psi} + 10 y_\W^2(1+y_{\sss \Psi}) + 3 y_\W^4) \over N_{HC} (1 + y_{\sss \Psi}^2 - y_\W^2) (4 - y_\Z^2)} \nonumber \\
&+& {4(4 + 20 y_\W^2 + 3 y_\W^4) \over N_{HC}^2 (4 - y_\Z^2)^2 } \Big]
\eea
\end{widetext}


\end{document}